\documentclass{aa}  

\usepackage{xcolor}
\usepackage{graphicx}
\usepackage{txfonts}
\usepackage{appendix}
\usepackage{lscape}
\usepackage[]{hyperref}

\begin{document}

   \title{The impact of AGN X-ray selection on the AGN halo occupation distribution}

   \author{M. C. Powell \inst{1,2}\thanks{Schwarzschild Fellow; e-mail: mpowell@aip.de}, M. Krumpe\inst{1}, A. Coil \inst{3} \and T. Miyaji\inst{4}}

\institute{ Leibniz-Institut fur Astrophysik Potsdam (AIP), An der Sternwarte 16, D-14482 Potsdam, Germany
\and
Kavli Institute for Particle Astrophysics and Cosmology, Stanford University, 452 Lomita Mall, Stanford, CA 94305, USA
\and
Department of Astronomy and Astrophysics, University of California, San Diego, 9500 Gilman Drive, La Jolla, CA 92093, USA 
\and
Universidad Nacional Autonomade Mexico( UNAM), Instituto de Astronomıa, AP106, Ensenada 22860, BC, M´exico
 }

   \date{Received ; accepted }

  \abstract
  {}{
   The connection between active galactic nuclei (AGN) and their host dark matter halos provides powerful insights into how supermassive black holes (SMBHs) grow and coevolve with their host galaxies. 
   Here we investigate the impact of observational AGN selection on the AGN halo occupation distribution (HOD) by forward-modeling AGN activity into cosmological N-body simulations.} {By assuming straightforward relationships between the SMBH mass, galaxy mass, and (sub)halo mass, as well as a uniform broken power law distribution of Eddington ratios, we find that luminosity-limited AGN samples result in biased HOD shapes.} {While AGN defined by an Eddington ratio threshold produce AGN fractions that are flat across halo mass (unbiased by definition), luminosity-limited AGN fractions peak around galaxy-group-sized halo masses and then decrease with increasing halo mass. With higher luminosities, the rise of the AGN fraction starts at higher halo masses, the peak is shifted towards higher halo masses, and the decline at higher halo masses is more rapid. These results are consistent with recent HOD constraints from AGN clustering measurements, which find (1) characteristic halo mass scales of $\log M_{Vir}\sim$ 12 - 13 [$h^{-1}M_{\odot}$] and (2) a shallower rise of the number of satellite AGN with increasing halo mass than for the overall galaxy population. 
   Thus 
   the observational biases due to AGN selection can naturally explain the constant, characteristic halo mass scale inferred from large-scale AGN clustering amplitudes over a range of redshifts,
   as well as the measured inconsistencies between AGN and galaxy HODs. }
    {We conclude that AGN selection biases can have significant impacts on the inferred AGN HOD, and can therefore lead to possible misinterpretations of how AGN populate dark matter halos and the AGN-host galaxy connection. }
   
   {}


   \maketitle
%
%________________________________________________________________

\section{Introduction} \label{sec:intro}
The connection between active galactic nuclei (AGN) and their cosmic web environments may provide powerful constraints on the mechanisms that trigger and quench supermassive black hole (SMBH) accretion and growth. There are several proposed mechanisms that trigger or quench AGN activity and depend on the galaxy's large-scale host dark matter halo environments. These include galaxy interactions \citep[e.g.,][]{Hopkins:2008}, 
ram pressure stripping \citep[e.g.,][]{Marshall:2018,Ricarte:2020}, and hot coronas in massive halos \citep[e.g.,][]{Bower:2017}. Understanding how AGN occupy dark matter halos can thus test these various scenarios for black hole growth.

In the hierarchical model of structure formation, dark matter halos are the peaks of the dark matter distribution that have gravitationally collapsed and grew by accreting mass and merging with other halos over time. Bound halos that have fallen into a larger halo are known as subhalos, and they are thought to host satellite galaxies. 
Those that are not within any larger virialized structure (and are sufficiently massive) host central galaxies at their centers and are known as `parent' halos.
Dark matter simulations have characterized the clustering statistics of dark matter halos across a wide range of cosmic time and mass scale. From this understanding, the clustering amplitudes of various observed galaxy populations can provide clues for how galaxies occupy their host dark matter halos \citep[e.g.,][]{Behroozi:2010,Zehavi:2011}.

The average number of galaxies that reside within a parent dark matter halo as a function of halo mass is known as the halo occupation distribution (HOD; \citealt{Berlind:2002,Cooray:2002}). This function can be disaggregated as a sum of two components: the central  galaxies (typically modelled as a 
 smoothed step function that is equal to zero at parent halo masses $M_{Vir} << M_{\rm threshold}$ and equal to unity $M_{Vir} >> M_{\rm threshold}$, where $M_{\rm threshold}$ is the threshold mass for a halo to be able to host a galaxy above a given stellar mass)
and satellite galaxies (modeled as a powerlaw 
 where the number of satellites increases with parent halo mass). The satellite power law is parameterized by its slope, $\alpha$, which determines how sharply the number of satellites hosted in a halo increases with mass. If $\alpha\sim 1$, then the number of objects within a parent halo scales linearly with parent halo mass. This is roughly the case for subhalos in dark matter simulations.

By analytically fitting the correlation function of galaxies from large, spectroscopic surveys, $\alpha$ has been found to be around unity for galaxy samples ranging in mass, luminosity, and color \citep[e.g.,][]{Zheng:2007,Zehavi:2011}. The HOD of AGN is much less certain due to their relatively low number density, which results in poorer clustering statistics. However, from AGN-galaxy cross correlation function measurements and group counts of AGN from X-ray surveys, $\alpha$ has been typically constrained to be less than unity \citep{Miyaji:2011,Allevato:2012,Krumpe:2015,Krumpe:2017, Powell:2018,Comparat:2023}.
This has been puzzling given the fact that a number of recent measurements have shown that AGN cluster like galaxies with matching parameters (stellar masses, redshifts, and sometimes star formation rates), inferring that AGN are statistically drawn from the overall galaxy population \citep[e.g.,][]{Mendez:2016,Yang:2018,Krishnan:2020,Aird:2021,Alam:2021}. In other words, if AGN activity stochastically occurs in all galaxies, then the AGN satellite slope is expected to be the same as for the galaxies (which is $\alpha \sim 1$). The fact that the $\alpha$ values have been found to be shallower in AGN HOD models have so far been interpreted as a physical process for black hole activity that depends on large-scale environment.    

Such a physical interpretation includes a scenario where AGN activity is suppressed in large group and cluster environments. This scenario is sensible since there is generally less available cold gas in cluster galaxies to fuel black hole accretion \cite[e.g.,][]{Catinella:2013}. It is also thought that major mergers are less efficient in cluster environments due to the high relative velocities of cluster members \citep[e.g.,][]{Hopkins:2008}, leading to fewer merger-triggered AGN in these systems. These ideas are consistent with studies of AGN in cluster environments \cite[e.g.,][]{Ehlert:2015,Noordeh:2020}.   
However, an issue in interpreting the shallow $\alpha$ values in terms of physical triggering mechanisms is that analytical HOD constraints often fail to fold in AGN selection effects and resulting biased samples. 

Recently, a simple empirical model connecting black holes, galaxies, and (sub)halos has been developed by populating AGN activity into halo catalogs from N-body simulations. By assuming straightforward relationships between the masses of black holes and galaxies, coupled with a distribution of Eddington Ratios, the clustering and luminosity function of a local sample of AGN has been reproduced after forward-modeling the hard X-ray-selection \citep{Powell:2022}. This work showed that a model in which the black hole mass correlates with both stellar mass {\it and} (sub)halo mass was preferred over a model without a black hole mass - (sub)halo mass correlation.
There were no assumed dependencies on environment or mass for the probability that a black hole accretes at a given Eddington rate; rather, each mock black hole was assigned an Eddington ratio drawn from the universal distribution constrained by AGN from the Swift/BAT Spectroscopic Survey (BASS; \citealt{Ananna:2022}).
This assumption was motivated by the findings from BASS and other X-ray surveys that the shape of this intrinsic distribution is mass-independent \citep{Aird:2012,Aird:2018,Georgakakis:2017,Ananna:2022}.

Apart from the possible physical interpretations for the different AGN and galaxy HODs, one key test has not yet been performed: whether or not observed differences are simply due to an observational bias arising from how the AGN are selected.
In this paper, we use AGN-(sub)halo models to investigate the impact of AGN selection on the resulting AGN HOD shape by forward-modeling Eddington- and  luminosity-limited AGN samples. We test whether or not the resulting satellite occupation slope is sensitive to the choice of selection, and we compare our results to previous $\alpha$ constraints from HOD modeling of AGN clustering measurements.

This paper is organized as follows: the empirical AGN-halo model and analysis method is described in Sections \ref{sec:model} and \ref{sec:mockhod}; the HOD results vs. selection technique is presented in Section \ref{sec:results}; we discuss our results and compare with previous AGN HOD constraints in Section \ref{sec:discussion}; we summarize our main conclusions in Section \ref{sec:summary}. We assume Planck 2015 cosmology (\citealt{Planck:2015}; $H_0=100\,h$ km s$^{-1}$ Mpc$^{-1}$, $h=0.677$, $\Omega_{m,0}=0.307$, $\Omega_{b,0}=0.0486$). $L_X$ always refers to the intrinsic, rest-frame $2-10$ keV luminosity.

\begin{figure*}
    \centering
    \includegraphics[width=.7\textwidth]{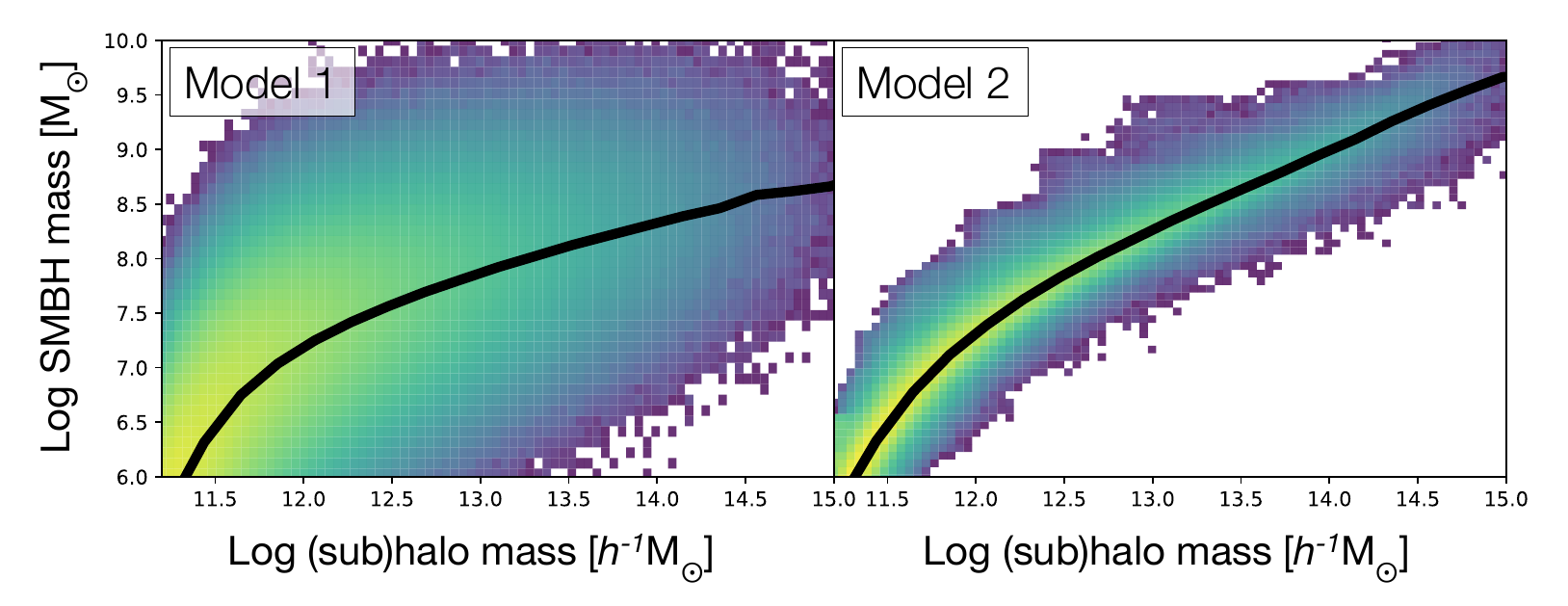}
    \caption{Distributions of black hole  mass  vs. (sub)halo mass (M$_{peak}$) according to the models assumed in this work (from \citealt{Powell:2022}), shown with a lognormal color scale. Model 1 assumes correlations between SMBH mass and stellar mass, as well as between stellar mass and (sub)halo mass. Model 2 includes an additional SMBH mass-(sub)halo mass correlation, leading to a much tighter relation between the two. }
    \label{fig:mbhmh}
\end{figure*}

\section{Empirical AGN-halo model} \label{sec:model}

We followed the method outlined in \cite{Powell:2022} to populate SMBHs into simulated dark matter halos and forward-model AGN samples. 
Because the SMBH masses and accretion rates are assigned probabilistically (described in detail below), we made many instances (or mock realizations) of AGN samples in order to account for the sample variance and to determine the uncertainties of the resulting AGN HODs.

We used the Unit N-body simulation \citep{Chuang:2019}, which is a $1~h^{-3}$Gpc$^{3}$ volume cube with a particle mass of $1.2\times 10^9~ h^{-1}$M$_{\odot}$ for the halo catalogs. This simulation assumed a Planck 2016 Cosmology and used the Rockstar halo-finder \citep{rockstar} to obtain the halo catalogs at each snapshot. We used the scale factor $a = 0.978$ ($z\sim 0.02$) snapshot to closely match to the redshift of the BASS AGN survey \citep{Koss:2022c,Powell:2022} that constrained the AGN-halo models; however, we repeated the analysis using other snapshots ($z=0.5, 1$), which produced similar results and are presented in the Appendix \ref{sec:appendix}.
We now outline the main steps in populating the simulation box with AGN.

\subsection{Populating halos with mock galaxies and SMBHs}

First, a mock galaxy was placed at the center of all halos and subhalos with peak virial masses $M_{\rm Vir}>5\times 10^{10}$ $h^{-1}$M$_{\odot}$\footnote{Throughout the paper we note $M_{\rm Vir}$ as the maximum virial mass of a subhalo throughout its lifetime (up until the time of the snapshot), rather than its instantaneous mass.}. A stellar mass was assigned to each mock galaxy according to the stellar mass-(sub)halo mass relation from \cite{Behroozi:2013}, which includes a log-normal scatter of 0.2 dex. 

Each mock galaxy was then populated with a black hole with a mass that depended either on (1) the stellar mass of the galaxy only (`Model 1') or (2) on the masses of both the galaxy and the (sub)halo (`Model 2'). These models each assume a powerlaw relationship between stellar mass ($M_{*}$) and black hole mass ($M_{BH}$) with a lognormal scatter. 

Because the galaxy-SMBH relation is uncertain (in terms of its normalization, slope, and scatter), we chose new sets of SMBH-galaxy parameters each time we generated a mock AGN realization. 
In this way, we marginalized over its uncertainties.
We drew from the constrained posterior distributions of the normalization, slope, and scatter from \citealt{Powell:2022}\footnote{The best-fit parameters of the $M_{BH}-M_{*}$ relation were a normalization of $7.76^{+0.24}_{-0.30}$ ($7.55^{+0.34}_{-0.39}$) dex [$M_{\odot}$], a scatter of $0.33^{+0.16}_{-0.18}$ ($0.43^{+0.16}_{-0.20}$) dex, and a powerlaw slope of $0.67^{+0.24}_{-0.22}$ ($0.51^{+0.21}_{-0.22}$) for Model 1 (2).} that produced population statistics consistent with the luminosity function and AGN-galaxy cross-correlation function at $z\sim 0.04$ (as calculated from the the Swift/BAT AGN Spectroscopic Surveys and 2MASS galaxy redshift survey \citep{Koss:2022c,Huchra:2012, Powell:2022}) for each mock realization. 
The masses were then assigned given the resulting relation, along with a random offset drawn from a lognormal distribution to account for the scatter.
For each model, we made many realizations to fully sample the posteriors.

In the case of the second model, there is an additional assumed correlation between SMBH mass and (sub)halo mass for fixed stellar mass. This was done via conditional abundance matching where, for fixed stellar mass, the most massive black holes are assumed to reside in the halos with the largest $M_{\rm Vir}$ and the least massive black holes in the smallest $M_{\rm Vir}$. This was done via the \texttt{conditional\_abunmatch} routine in the \texttt{halotools} Python package \citep{Hearin:2017}, which preserves the $M_{BH}-M_{*}$ and $M_{*}-M_{Vir}$ relations while introducing the third $M_{BH}-M_{Vir}$ correlation 
(more details are given in \citealt{Powell:2022}). Examples of two $M_{BH}-M_{\rm Vir}$ relations for each model are shown in Figure \ref{fig:mbhmh}. 

\citealt{Powell:2022} found that Model 2 was preferred by the Swift/BAT AGN Spectroscopic dataset by $2-5\sigma$ (where the precise significance depended on whether or not an independent constraint on $M_{BH}-M_{vir}$ using individual, nearby galaxies was used). However, we performed the analysis for both models in this work.

\subsection{Assigning accretion rates to the mock SMBHs}
The mock SMBHs were then assigned Eddington ratios ($\lambda_{Edd}\equiv L_{bol}/L_{Edd}$, where $L_{Edd}$ is the luminosity at which
radiation from the accretion disk balances gravity and is equal to $1.26\times 10^{38}(M_{\rm BH}/M_{\odot})$) by probabilistically drawing from a universal Eddington Ratio Distribution Function (ERDF). We assumed the measured distribution from (\citealt{Ananna:2022}; ERDF 1), which has the following broken powerlaw form:
\begin{equation}
    \frac{dN}{d\log \lambda_{Edd}} \propto \xi_{*} \times \left[ \left( \frac{\lambda_{Edd}}{\lambda_{*}} \right)^{\delta_1} + \left( \frac{\lambda_{Edd}}{\lambda_{*}} \right)^{\delta_2}\right]^{-1} .
\end{equation}

\noindent This study constrained these ERDF parameters ($\xi_*$, $\lambda_*$, $\delta_1$, $\delta_2$)\footnote{$\log \xi_*=-3.6$; $\log \lambda_*=-1.338$; $\delta_1=0.38$; $\delta_2=2.26$} by forward-modeling the distributions of the AGN luminosities, black hole masses, and Eddington ratios in the BASS survey. 

The use of a universal Eddington ratio distribution corresponds to the assumption that AGN are triggered by the same secular processes across mass scale and environment (i.e., no environmental factors). 
This is consistent with findings from constraints of the ERDF from various X-ray surveys \cite[e.g.][]{Aird:2012,Georgakakis:2017,Ananna:2022,Birchall:2022}, as well previous AGN clustering measurements that showed no significant trends with Eddington ratio \citep[e.g.,][]{Krumpe:2015,Krumpe:2023,Powell:2018}. 
However, some work has shown evidence that there is a dependence on galaxy mass, where more massive galaxies have a higher probability to host AGN (i.e., that the normalization of the ERDF is mass-dependent; \citealt{Aird:2018}). We therefore additionally tested an ERDF with the same shape as before, but where the normalization of the Eddington ratio distribution varied by an order of magnitude over the stellar mass range $10-11.5$ [dex $h^{-1}M_{\odot}$], which we label as `ERDF 2'. In this case, more massive galaxies were more likely to host AGN than less-massive galaxies.

\subsection{Selecting mock AGN}
With black hole masses and Eddington ratios assigned in each mock galaxy, the AGN bolometric luminosities ($L_{bol}$) were then inferred via $L_{bol} = 1.26\times 10^{38}(M_{\rm BH}/M_{\odot})\lambda_{\rm Edd}$ s$^{-1}$erg.
X-ray luminosities were estimated by assuming a constant bolometric correction ($\kappa$) of 20 (i.e., $L_{bol}=20\times L_{X}$; \citealt{Vasudevan:2007}). AGN were then selected based on either an X-ray luminosity or Eddington ratio threshold.
As an additional test, we also performed the analysis assuming an Eddington-ratio-dependent bolometric correction from \citep{Vasudevan:2007}, which we label as $\kappa(\lambda_{Edd})$.

Finally, to estimate uncertainties in the resulting AGN HODs, 
we made $\sim 50$ AGN mock realizations for each AGN-halo model. The HOD parameters for each AGN selection were examined via their distributions over the many realizations. This process is described in more detail in the following section.

\begin{figure*}
    \centering
    \includegraphics[width=.49\textwidth]{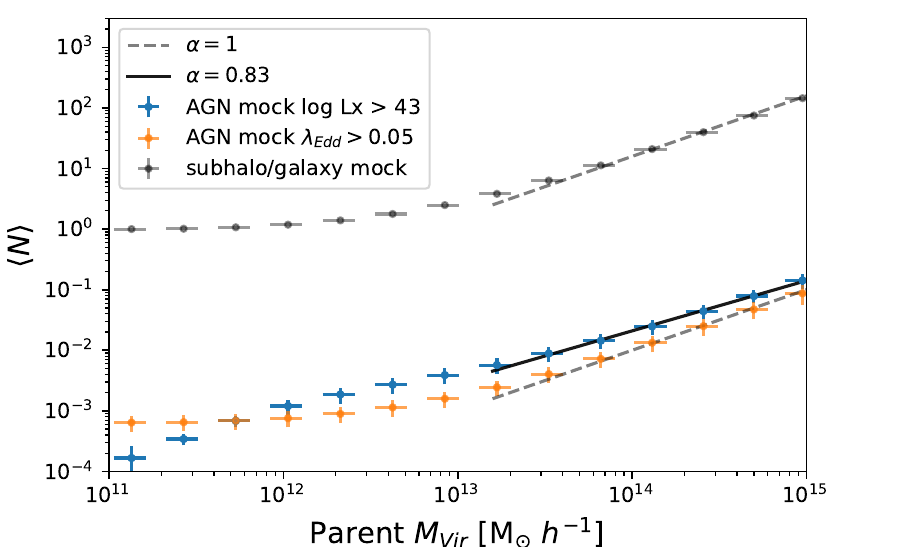}
    \includegraphics[width=.49\textwidth]{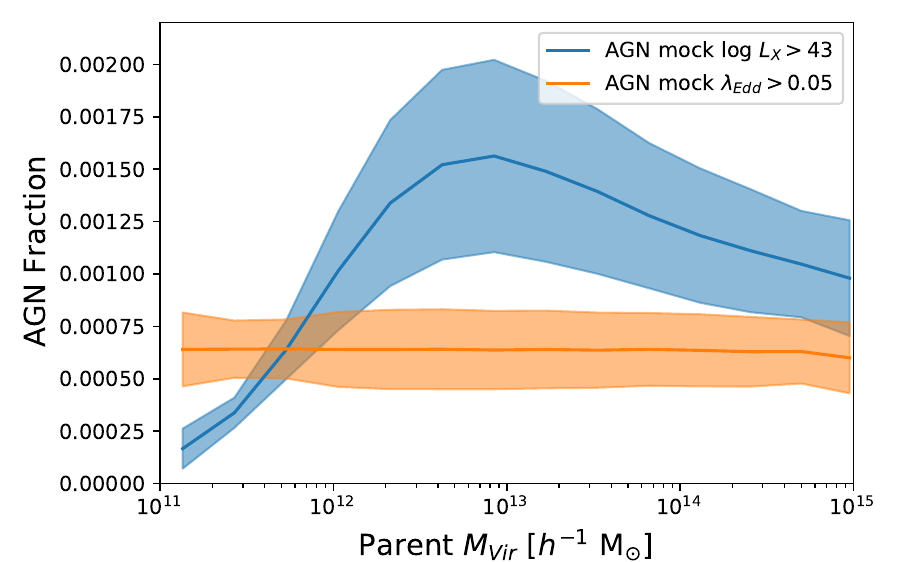}
    \caption{The impact of different AGN definitions on their host dark matter halo distributions. {\it Left:} HODs of mock AGN in our Model 2 simulation framework selected by an Eddington ratio threshold (orange) and a luminosity threshold (blue). Different slopes are visible above 10$^{13}$ for the two different AGN selection methods. The HOD of the full subhalo/galaxy population is shown by the gray data points for comparison which is identical in shape to the Eddington-limited AGN sample. {\it Right:} fraction of mock galaxies classified as AGN by the same definitions as in the left plot. The fraction remains constant (unbiased) with halo mass for Eddington-limited AGN (by definition); however, the luminosity-limited sample has a strongly biased halo occupation distribution due to their selection.}
    \label{fig:lxvsedd}
\end{figure*}

\section{Mock AGN HOD calculation}
\label{sec:mockhod}
The AGN halo occupation distribution of each mock realization was calculated by averaging the number of mock AGN residing within each bin of parent virial halo mass. The bins were defined as 15 logarithmic bins from $10^{11}< M_{vir}<10^{15}~h^{-1}M_{\odot}$. 
For comparison, we also calculated the subhalo (i.e., galaxy) HOD by the same method; namely, by averaging the number of subhalos with virial masses $M_{vir}>10^{11}$ [$h^{-1}M_{\odot}$] hosted 
within the same bins of parent halo mass.
Because high-Eddington accretion events are much rarer than low-Eddington accretion, most parent halos within the simulation box did not host a detectable mock AGN (by typical luminosity or Eddington ratio thresholds). Therefore, the AGN HOD has a much smaller normalization than the galaxy HOD. 
We also determined the AGN fraction vs. parent halo mass by dividing the AGN HOD by the overall subhalo HOD.

Observationally, X-ray AGN are detected from flux-limited surveys and are typically defined based on their luminosities.
For each AGN-halo model, we computed the AGN HOD for several X-ray luminosity limits:
$10^{40}$, $10^{41}$, $10^{42}$, $10^{43}$ and $10^{44}$ erg s$^{-1}$. This was done for each $M_{BH}-M_{vir}$ model (Model 1 and 2) coupled with each ERDF (ERDF 1 and ERDF 2). We also tested each Models 1 and 2 with the Eddington-ratio-dependent bolometric correction and ERDF 1 (see Table \ref{tab:alphas} for a summary of these results). These luminosity-limited AGN HODs were also compared to those of AGN defined by an Eddington ratio threshold, which corresponds to an unbiased sample (since by definition the Eddington ratios were independent of environment).

The satellite slope ($\alpha$) was estimated by fitting a powerlaw to each HOD realization for $\log\,M_{vir}>13$ [$h^{-1}$ M$_{\odot}$] using least-squares minimization via the {\tt curve\_fit scipy} routine, weighted by the square root of the number of AGN per bin. A threshold of $\log\,M_{vir}=13$ [$h^{-1}$ M$_{\odot}$] was chosen because satellite galaxies dominate the HOD above masses of $\log\,M_{vir}>13$ [$h^{-1}$ M$_{\odot}$] based on previous measurements of typical $L_*$ galaxies \citep[e.g.,][]{Zehavi:2011}. 
The uncertainties in the satellite slopes ($\alpha$) for each model and AGN definition were calculated by the 16th and 84th percentiles of the measurements across all AGN mock realizations.

\section{Results} \label{sec:results}

\begin{figure}
    \centering
    \includegraphics[width=.49\textwidth]{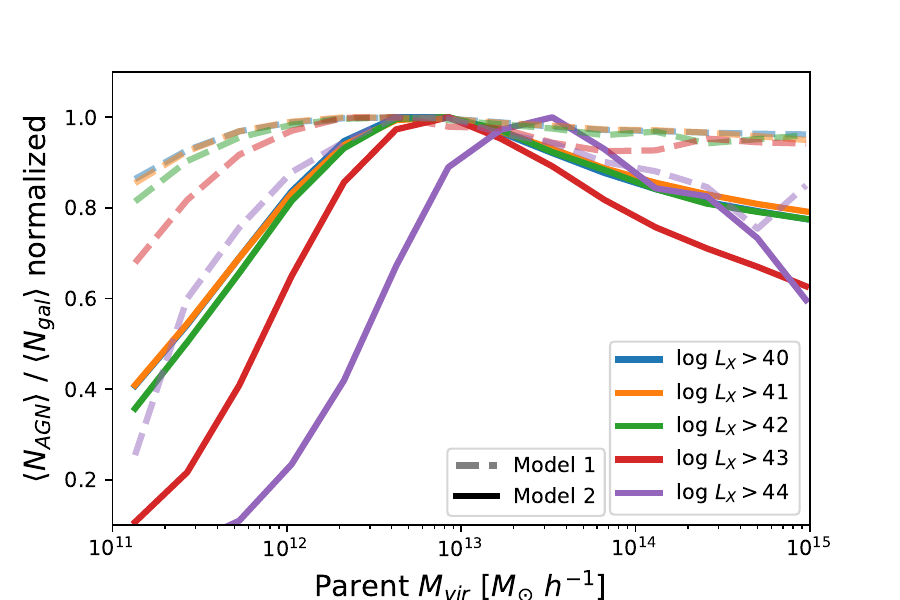}
    \caption{Normalized AGN fractions 
    (averaged over all realizations) vs. parent halo mass. Each luminosity threshold is shown in different colors, and the dashed and solid lines correspond to Model 1 and Model 2, respectively. Higher thresholds correspond to AGN fractions that are more peaked, especially for Model 2.}
    \label{fig:fagns}
\end{figure}

In this section we examine the HODs of mock AGN selected by a luminosity threshold vs. by an Eddington ratio threshold. We also examine trends with luminosity threshold and report the resulting $\alpha$ values under each AGN-halo modeling assumption.

\subsection{Luminosity-limited vs. Eddington-limited AGN selection} \label{sec:selections}

The HODs of subhalos, AGN with log $L_X > 43$ [erg s$^{-1}$] (i.e., $L_{bol}>44.3$ [erg s$^{-1}$]), and AGN with log $\lambda_{Edd} > 0.05$ within our simulation framework (assuming Model 2 and ERDF1) is shown by the the left-hand panel of Fig. \ref{fig:lxvsedd}. The AGN fraction vs. halo mass  (i.e., the AGN HOD divided by the galaxy HOD) for both definitions is shown by the right-hand panel, demonstrating 
how different AGN selections based on the identical parent AGN sample yield significantly different HOD shapes.
While the HOD of the Eddington-limited AGN only differs from the overall subhalo HOD by its normalization, corresponding to a flat AGN fraction across halo mass, the luminosity-limited AGN sample has a very different HOD shape. Note that each AGN definition draws from the same sample of mock accreting black holes; therefore, this HOD shape disparity is purely due the luminosity selection.

At lower parent halo masses, AGN are more likely to be missed by X-ray surveys due to their typically smaller-mass black holes and resulting fainter AGN luminosities. This leads to a drop off in AGN fraction at $M_{vir}<10^{13}$ $h^{-1}$M$_{\odot}$. The AGN fraction peaks at $M_{vir}\sim 10^{13}$ $h^{-1}$M$_{\odot}$ 
and again decreases with increasing halo mass at $M_{vir}\sim 10^{13}$. This corresponds to a satellite slope ($\alpha$) less than one, and is explained again by the selection effect against lower-mass black holes; as the parent halo mass increases, the more satellite galaxies it tends to host. AGN in satellites are more likely to be missed by surveys because they are generally in lower mass systems with lower mass black holes. This mass selection bias is discussed in more detail in Section \ref{sec:discussion}.

\begin{figure*}
    \centering
    \includegraphics{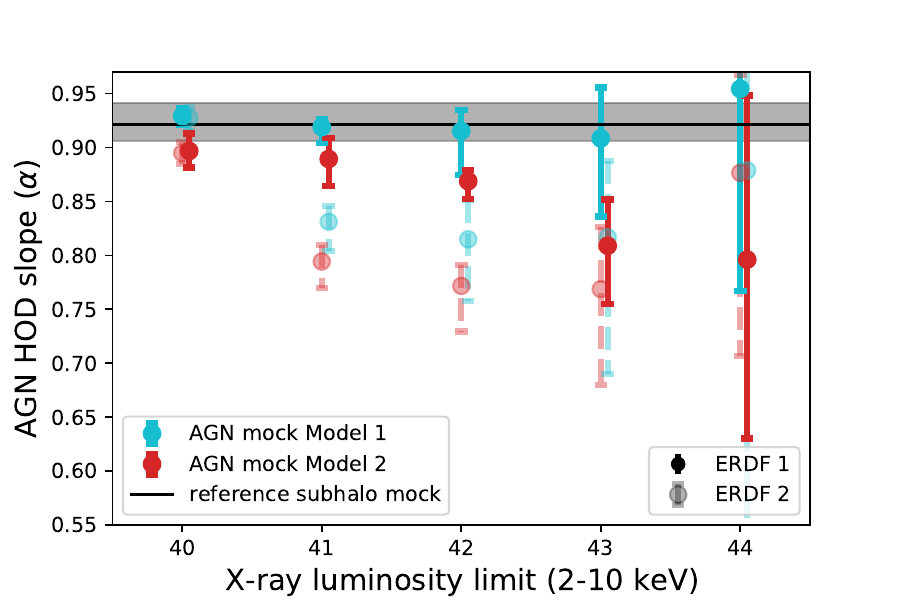}
    \caption{Range of intrinsic AGN HOD slopes ($\alpha$) for parent halos with $\log~M_{vir}>13$ [$h^{-1}$M$_{\odot}$] as a function of X-ray luminosity limit. The two colors correspond to each $M_{\rm BH}-M_{\rm vir}$ model and their accepted parameters. 
    The solid data points correspond to the standard ERDF 1, while the transparent, dashed data points correspond to ERDF 2 (which includes a strong stellar mass dependence).
    The slope for all subhalos is shown in gray for comparison.  
    In the case of Model 2, where black hole mass correlates more strongly with halo mass, the resulting median AGN HOD slope decreases and becomes more biased against the overall population of accreting black holes for higher luminosity thresholds. 
    For both models, ERDF 2 produces smaller values of alpha, especially for the $L_X=10^{41}$ and $10^{42}$ erg s$^{-1}$ thresholds.
    }
    \label{fig:avL}
\end{figure*}

\subsection{Trends with luminosity threshold} 
We repeated the investigation for other luminosity thresholds and found similar shapes for the AGN fraction vs. halo mass. To show how the shape of each differs for each threshold, we plot
their normalized AGN fractions in Fig. \ref{fig:fagns} for both $M_{BH}-M_{vir}$ models. As the luminosity threshold increases, the fractions tend to become more and more biased over the parent AGN sample, and the peak is shifted to higher halo mass scales. This is especially the case for Model 2.
The shapes of these AGN fractions with parent halo mass for Model 1 agree qualitatively with what was found in \citealt{Jones:2017}, which also used a semi-numerical prescription for AGN accretion within a dark matter simulation.

We plot the satellite slope trends for both models in Fig. \ref{fig:avL}, where the median $\alpha$ values calculated for each AGN definition (via $\sim 50$ model realizations) are compared to the satellite slope calculated for the full sample of subhalos within the simulation. These values are also given in Table \ref{tab:alphas} for each modeling assumption.
The size of the errorbars on the median alpha values are due to (1) the range of allowed $M_{BH}-M_{*}$ parameters that each realization draws from and (2) the limited number statistics in the simulation box.

We found that, for Model 2, $\alpha$ tends to decreases with increasing AGN luminosity threshold, meaning that the HOD shape becomes more biased against the overall population of accreting black holes as more 'typical' AGN are missed by the selection. 
For both models the uncertainties increase towards larger $L_X$ limit due to the decreasing number of objects in the simulation with higher luminosities.
However, the trends between $\alpha$ and the luminosity limit are different for Model 1.
For Model 1, no significant differences between the median  $\alpha$ values of the AGN and the underlying population as a function of $L_X$ is found. 
This means that the systematic selection bias in the HOD shape depends not only on the AGN luminosity limit, but also on the underlying physical relationship between SMBH mass and (sub)halo mass. While this relation has yet to be firmly constrained,
Model 2, in which the SMBH mass more tightly correlates with its host subhalo, has been found to be preferred by local datasets \citep{Marasco:2021,Powell:2022}.

Shallow $\alpha$ values were also found when assuming both Eddington ratio distribution functions: the universal distribution (ERDF 1), and the one whose normalization varied with stellar mass (ERDF 2). The alpha values found for both models and ERDFs (for each luminosity threshold) are given in Table \ref{tab:alphas}. The mass-dependent ERDF (ERDF 2) produced slightly shallower AGN satellite slopes for moderate luminosity cuts (i.e., 0.77 vs. 0.81 for log $L_X > 43$ [erg s$^{-1}$]) for both models (Fig. \ref{fig:avL}). This is because the additional ERDF dependence on stellar mass (and therefore on (sub)halo mass) further decreased the probability for AGN in lower mass galaxies/(sub)halos to be detected. Therefore, any mass dependence of the Eddington ratio distribution function may also contribute smaller measurements of $\alpha$.
Lastly, we find no significant differences in the results or trends found between the constant bolometric correction and the Eddington ratio-dependent case.

\begin{table*}
    \centering
    \begin{tabular}{|c|cccc|}
        \multicolumn{5}{c}{ {\bf $\alpha$ values} ($z=0$)} \\
         \hline
         \hline
         Log $L_X$ limit [erg s$^{-1}$] &  41  & 42 &   43 & 44  \\
         \hline
         Model 1; ERDF 1 & $0.92\pm0.01$ & $0.92\pm0.02$ & $0.91\pm0.05$ & $0.94\pm0.19$ \\
        Model 1; ERDF 2 & $0.83\pm0.02 $ & $0.82\pm0.05$ & $0.82\pm0.10$ & $0.88\pm0.35$  \\
         Model 2; ERDF 1 & $0.89\pm0.02$ & $0.87\pm0.02$ & $0.81\pm0.05$ & $0.80\pm0.16$ \\
        Model 2; ERDF 2 & $0.79\pm0.02$  & $0.77\pm0.03$ & $0.77\pm0.08$ & $0.87\pm0.17$\\
        Model 1; ERDF 1; $\kappa(\lambda_{Edd})$& $0.90\pm0.01$ & $0.90\pm0.02$ & $0.87\pm0.05$ & $1.00\pm0.27$\\
        Model 2; ERDF 1; $\kappa(\lambda_{Edd})$ & $0.87\pm0.01$ & $0.86\pm0.02$ & $0.81\pm0.04$ & $0.85\pm0.16$  \\

         \hline
    \end{tabular}
    \caption{Satellite slopes ($\alpha$ values) for the different $L_X$ thresholds and modeling assumptions in our simulation framework. In addition to the two $M_{BH}-M_{Vir}$ models (Model 1 and Model2), we tested two different Eddington ratio distribution functions; one universal distribution (ERDF1; \citealt{Ananna:2022}) and one whose normalization is mass-dependent (ERDF2). Lastly, while a constant X-ray bolometric correction was assumed  by default, we also calculated $\alpha$ for each model assuming an Eddington-ratio-dependent BC ($\kappa(\lambda_{Edd})$; \citealt{Vasudevan:2007}).}
    \label{tab:alphas}
\end{table*}

While this analysis focused on the low-redshift regime from using the BASS survey constraints, similar overall trends were found with higher-$z$ simulation snapshots ($z\sim 0.5$ and $z\sim 1$) under the assumption that the same AGN-halo connection as found for low-redshift can be applied to higher redshifts (see Appendix \ref{sec:appendix}). However, we note that the relationships between $M_{\rm BH}$ and $\lambda_{\rm Edd}$ with environment are even less understood at moderate-to-high redshifts and may be different than the local Universe.

\begin{figure*}
    \centering
    \includegraphics[width=.45\textwidth]{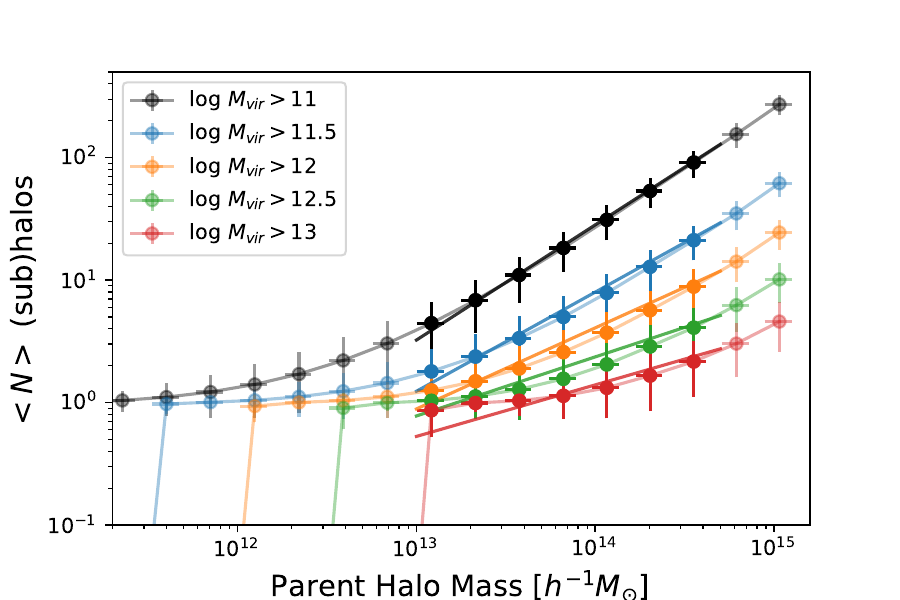}
    \includegraphics[width=.45\textwidth]{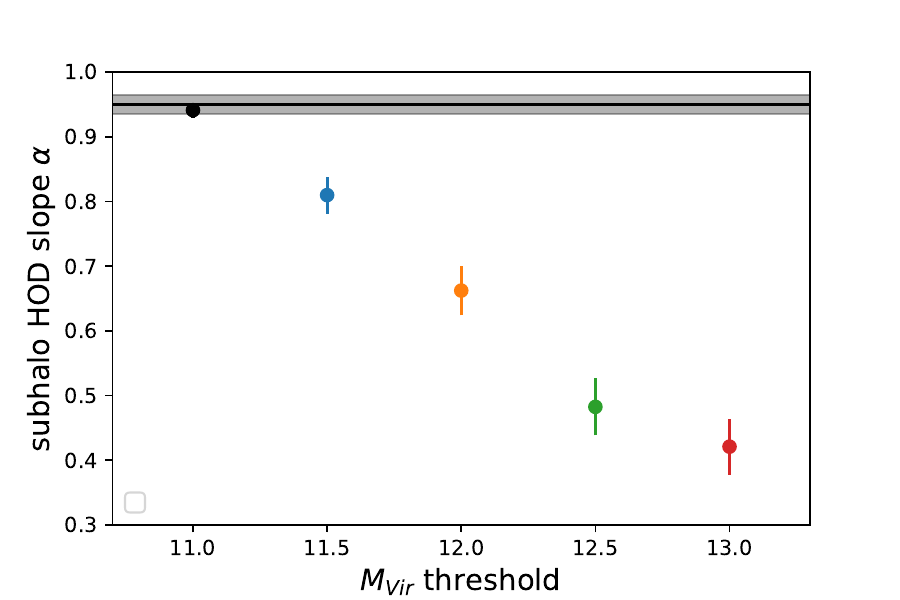}
    \caption{Subhalo HOD shapes for various mass thresholds. {\it Left:} Average number of subhalos in a parent halo as a function of parent halo mass for several peak virial (sub)halo mass limits. The solid colored lines correspond to the fitted powerlaw in the parent halo mass range $10^{13}< M_{Vir} < 5\times 10^{14}$ $h^{-1}M_{\odot}$. {\it Right:} HOD slopes ($\alpha$) of each subsample of subhalos, showing the apparent decline of $\alpha$ with mass threshold.}
    \label{fig:avsmv}
\end{figure*}

\section{Discussion}
\label{sec:discussion}

\subsection{Interpreting the shallow AGN satellite slopes} 

X-ray detection provides one of the most efficient AGN selection methods. However, since $L_X \propto \lambda_{Edd} \times M_{\rm BH}$, this selection technique is biased against low-mass SMBHs; unless they are accreting at very high Eddington ratios, smaller-mass black holes are likely to be undetected since they are less luminous. At the same time, SMBHs accreting at low Eddington ratios are also missed in X-ray surveys, and this is especially the case in lower-mass systems. Because more massive galaxies tend to have more massive SMBHs (and since X-ray emission approximates the total accretion luminosity), X-ray AGN activity tends to be more difficult to detect in smaller-mass galaxies \citep[e.g.,][]{Aird:2012}.

In the hierarchical model of structure formation, dark matter halos grow by accreting matter and merging with each other. Because of this, larger-mass halos tend to have more substructure (and host more satellite galaxies) than smaller mass halos due to their many mergers over cosmic time. A bias against AGN in low-mass (sub)halos would cause {AGN to more likely be undetected in (1) central galaxies in lower-mass parent halos and (2) satellites, since each of these have smaller masses and tend to host less luminous AGN. Therefore, imposing a flux or luminosity limit results in 
an AGN fraction that initially rises with halo mass (as AGN in central galaxies become easier to detect), peaks around group-size halo masses, and then declines 
with halo mass in the regime where the number of satellite galaxies start dominating the HOD ($M_{vir} \sim 10^{13}~h^{-1}M_{\odot}$), as we show in Figs. \ref{fig:fagns} and \ref{fig:avL}. 
The AGN fraction decreases in the massive halo regime 
because AGN are harder to detect in the lower-mass satellite galaxies due to hosting smaller-mass SMBHs, and this effect
corresponds to a shallower HOD power-law slope than for inactive galaxies (e.g., $\alpha<1$). It is purely due to an imposed AGN selection based on a luminosity  threshold.

These shallow AGN satellite slopes (and the AGN fraction decline at high halo masses) can alternatively be understood by the following: more massive satellite galaxies are more likely to be hosted in more massive parent halos, as opposed to less massive parent halos. Consequently, the HOD of subhalos moves rightward to higher parent halos masses for higher and higher (sub)halo mass thresholds. This is shown in Fig. \ref{fig:avsmv} (left), where the average number of subhalos is plotted as a function of parent halo mass for several (sub)halo virial mass thresholds; each HOD has the same shape and slope, but begin at different parent halo mass scales. When fitting the HODs with a simple powerlaw (as is often done in observational clustering studies) within the same given parent halo mass range, this results in comparatively shallower satellite slopes ($\alpha$ values) for increasing mass thresholds (Fig. \ref{fig:avsmv}, right). We limited the halo mass range $M_{\rm Vir}<5\times 10^{14}~h^{-1}$M$_{\odot}$ for this calculation since the volumes of most X-ray surveys are too small to statistically probe AGN within these rare, highest halo masses. Therefore, since X-ray AGN samples are biased against small-mass SMBHs (and therefore against small subhalo masses), then the AGN HOD is likely to have a shallower slope than galaxy samples, since observed galaxies are generally more complete towards lower stellar/(sub)halo masses than AGN samples.

The selection effect against AGN in lower-mass (sub)halos is more significant when there is a tighter correlation between SMBH mass and (sub)halo mass (i.e., Model 2), since there is also a stronger trend between (sub)halo mass and AGN luminosity. 
We show this in Fig. \ref{fig:LxvsMh}, which plots the mock AGN X-ray luminosity vs. (sub)halo mass for each $M_{BH}-M_{*}$ Model, where the colors correspond to their Eddington ratios (assuming ERDF 1 and limiting the mocks to $\log~\lambda_{Edd}>-4$). Model 2 results in a tighter correlation between $L_X$ and (sub)halo mass. Therefore, 
when enforcing a luminosity limit, it is more probable that the AGN in lower-mass systems are missed in Model 2 then in Model 1. This is why the trends shown in Figures \ref{fig:lxvsedd} and \ref{fig:fagns} are different for each model; the larger scatter between AGN luminosity and (sub)halo mass in Model 1 blurs out the mass dependence of the AGN detection probability, and leads to little or no trend between $\alpha$ and a luminosity limit. Consequently, measurements of shallow $\alpha$ values from previous AGN clustering studies may provide further evidence that Model 2, where SMBH mass tightly correlates with (sub)halo mass, is more likely to be true than Model 1. 

\begin{figure*}
    \centering
    \includegraphics[width=.9\textwidth]{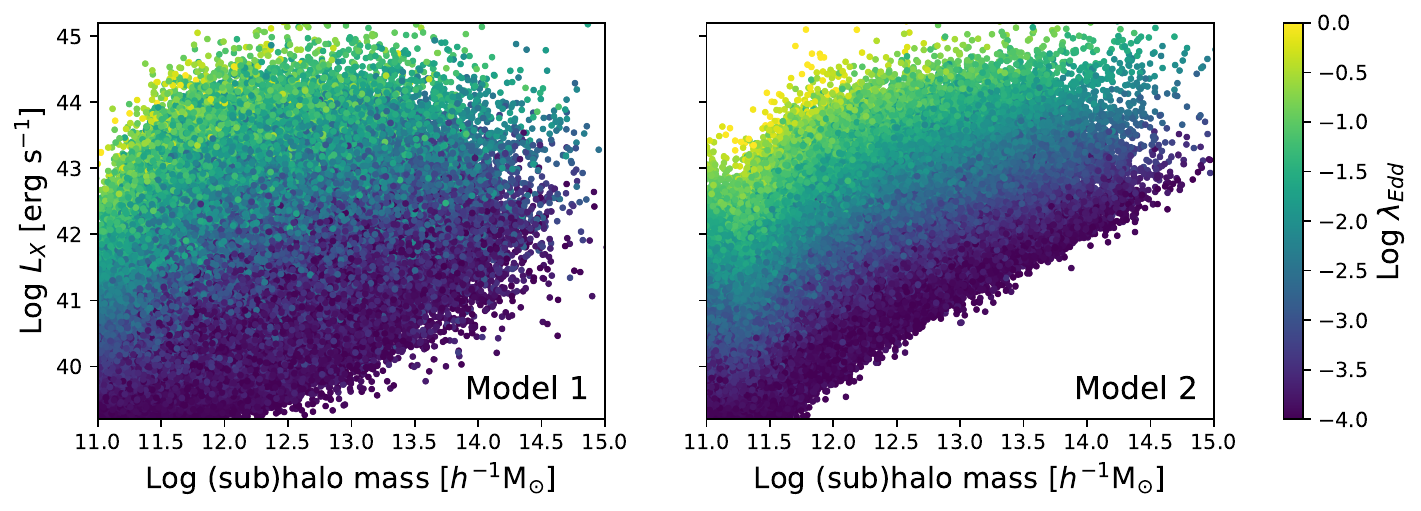}
    \caption{X-ray Luminosity vs. host (sub)halo mass of mock AGN with $\log~\lambda_{Edd}>-4$ for Model 1 (left) and Model 2 (right). The color of each data point corresponds to Eddington ratio. Model 2 results in mock AGN that have a tighter correlation between luminosity and (sub)halo mass.}
    \label{fig:LxvsMh}
\end{figure*}

The peak AGN fraction at log $M_{vir} = 12.5-13$ [$h^{-1}$M$_{\odot}$] is consistent with typical host halo mass estimates obtained from the large-scale (i.e. $1-10$ $h^{-1}$Mpc) clustering amplitudes of X-ray AGN over a wide range of luminosity and redshift \citep[e.g.,][]{Cappelluti:2012,Powell:2020}. This work shows that this characteristic halo mass scale could be purely caused by an observational bias due to the fact that the probability for finding an X-ray AGN peaks at this parent halo mass when assuming typical X-ray luminosity limits; Fig. \ref{fig:lxvsedd}, right). Indeed, \citealt{Aird:2021} found that when incorporating the probabilities for AGN activity within simulated galaxies, the average parent halo masses of $L_X$-limited AGN were also around this characteristic mass scale.
This highlights the importance of folding AGN selection in to clustering analyses and their interpretations; before the details of how selection in observed AGN samples impacts the AGN HOD is understood, any physical interpretation of the HOD shape has limited scientific value.

\subsection{Comparison to previous measurements} \label{sec:prev}

While AGN HOD measurements have been limited due to current survey statistics, cross-correlations between AGN and galaxies have provided HOD constraints for several X-ray and optical AGN surveys.
First, \citealt{Miyaji:2011} cross-correlated AGN from the \textit{ROSAT} All-Sky Survey (RASS) with the SDSS LRG sample, and by fitting the cross-correlation function analytically, constrained $\alpha<1$. This was recently confirmed with the updated samples of the RASS/SDSS DR14 broad-line AGN cross-correlated with the SDSS CMASS galaxy sample in the redshift range $0.44<z<0.64$ \citep{Krumpe:2023}. Their sample has median luminosity X-ray of $\sim 44.8$ [erg s$^{-1}$], and so this is consistent with our results for Model 2 and/or ERDF 2.
At similar redshifts ($z\sim 0.35$), \citealt{Comparat:2023} found that the best-fit HOD model for AGN in eFEDS had $\alpha = 0.73\pm 0.38$, in agreement with these previous findings.

In deeper, smaller-volume X-ray surveys, uncertainties on the AGN HOD have been larger \citep[e.g.,][]{Richardson:2013}. However, by directly counting the number of X-ray-selected AGN within X-ray-selected galaxy groups and clusters in the COSMOS survey at $0<z<1$, \cite{Allevato:2012} determined $\alpha<0.6$. This is quite low; however, this could be because the smaller-volume is less sensitive to the rare, higher-mass systems where the satellites dominate the HOD.
On the other hand, \cite{Leauthaud:2015} inferred $\alpha\sim 1$ for AGN in COSMOS via weak lensing measurements. In their analysis, they showed that the AGN lensing signal was consistent with galaxies with similar stellar masses and redshifts as the data, thereby taking the incompleteness of the survey into account and the effects of AGN selection biases.

In the local Universe, cross-correlations between hard X-ray selected AGN from Swift/BAT with 2MASS galaxies at $0.01<z<0.1$ resulted in HOD constraints with $\alpha=0.8^{+0.2}_{-0.5}$ \citep{Powell:2018}. \citealt{Krumpe:2017} did a similar analysis with Swift/BAT and INTEGRAL/IBIS AGN at $0.007<z<0.037$ and found $0.68 < \alpha<1$, with the galaxy sample having $\alpha=1.13\pm0.07$. 
Moreover, \cite{Powell:2018} found that the clustering measurement matched predictions based on the stellar mass distribution of the AGN, again showing no difference between AGN and galaxy halo occupation when taking the survey incompleteness and AGN selection biases into account.

These previous X-ray measurements favoring shallow $\alpha$ values (and typical halo masses of $\sim 10^{13}~h^{-1}M_{\odot}$) may be further evidence that Model 2, where SMBH mass tightly correlates with (sub)halo mass, is more likely to be true than Model 1, since Model 2 predicted shallower alpha values. However, further investigation into the environmental dependence of AGN activity vs. redshift is needed, as well as careful consideration of the modeling assumptions and the observational biases that go into measuring and interpreting AGN HODs.

For optically-selected AGN, the satellite power-law slope has tended to be more consistent with galaxies \citep{Richardson:2012, Chatterjee:2012,Shen:2013,Krumpe:2023}. 
Optical selection, which usually involves color cuts and/or emission line diagnostics, comes with a distinct set of selection biases that are very different from X-ray selections and depend on the host galaxy properties (i.e., stellar mass, star formation rate, and dust content; \citealt{azadi:2017}). Therefore, the resulting HOD selection biases are less straightfoward to model and to interpret than an X-ray luminosity threshold for an X-ray selected AGN sample. Future work forward-modeling multiple AGN selection methods are needed to determine whether optical quasars occupy their host dark matter halos consistently with X-ray AGN.

\section{Summary and conclusion}
\label{sec:summary}

We have investigated the impact of X-ray AGN selection on measurements of the halo occupation distribution by forward-modeling AGN activity into halo catalogs from N-body simulations. By assuming correlations between the black hole mass, stellar mass, and subhalo mass from \cite{Powell:2022} coupled with a universal Eddington Ratio distribution function, we measured the AGN HOD for several different AGN definitions based on Eddington ratio and luminosity thresholds to investigate the sensitivity of the resulting measured satellite slope. Our main conclusions are as follows:

\begin{itemize}
    \item X-ray luminosity-limited AGN selection can significantly impact the resulting AGN fraction as a function of halo mass, leading to biased measurements of the AGN halo occupation distribution. 
    \item We find that for AGN samples with typically-used X-ray luminosity thresholds, the AGN fraction peaks at $\sim 10^{13}~h^{-1}$ M$_{\odot}$ and decreases with halo mass. This corresponds to a satellite HOD slope $\alpha < 1$, which agrees with estimates from AGN clustering measurements (and differs for galaxies, where $\alpha\sim 1$).
    \item For the case where black hole mass correlates with (sub)halo mass, the resulting AGN satellite power-law slope $\alpha$ decreases with increasing AGN luminosity thresholds.

\end{itemize}

The findings presented in this paper are the result of luminosity X-ray selection alone, and not from any assumed trends between SMBH accretion physics and environment. It is  notable that we are able to reproduce previously found HOD results without invoking such triggering dependencies. 

To conclude, AGN selection methods can result in missed accreting SMBHs in certain galaxy and/or AGN types. The missed AGN may correlate with the mass scale of their host halos, leading to a warped HOD. 
With several future surveys on the horizon that will provide order-of-magnitude better clustering statistics (e.g., eROSITA, \textit{Athena}), understanding AGN selection effects will become even more crucial for characterizing the underlying AGN-halo connection and its evolution. 
If we do not correctly account for these observational biases, we cannot claim to understand the physics that gives rise to the shape of the AGN HOD.
Forward-modeling the AGN population in simulations provides a promising way to interpret these future high-precision measurements correctly while having full control of the selection biases.

\begin{acknowledgements}
     This research made use of Astropy, a community-developed core Python package for Astronomy \citep{2018AJ....156..123A, 2013A&A...558A..33A}, halotools v0.8, specialized Python package for building and testing models of the galaxy-halo connection \citep{Hearin:2017}, matplotlib, a Python library for publication quality graphics \citep{Hunter:2007}, NumPy \citep{harris2020array}, and SciPy \citep{Virtanen_2020}. 
     M.K. acknowledges support by DLR grant FKZ 50 OR 2307. TM is supported by UNAM-DGAPA IN 114423.   
\end{acknowledgements}

\bibliography{references}
\bibliographystyle{aa}

\begin{appendix}

\section{Higher-z trends}
\label{sec:appendix}

We repeated the analysis at higher redshift snapshots by extrapolating Model 1 and Model 2 to $z=0.5$ and $z=1$. We used the galaxy-halo models at the respective redshifts \citep{Behroozi:2010} with the same galaxy-SMBH relation as before for $z\sim 0.04$. The same galaxy-SMBH relation was used because the $M_{BH}-M_{*}$ relation has not been found to significantly evolve with redshift \citep[e.g.,][]{Suh:2019}. The parameters of the universal ERDF were modified such that resulting mock X-ray luminosity function were consistent with the $z=0.5$ and $z=1$ XLF from \cite{Gilli:2007} and \cite{Marchesi:2020}. Fig. \ref{fig:highz} shows the $\alpha$ values vs. luminosity limit from the mocks generated by each model in these higher-z snapshots. Similar trends are seen as at low-z, although the error bars are larger due to the poorer numbers of massive systems at these cosmic epochs (therefore making it harder to constrain $\alpha$). The normalized AGN fractions are plotted in Fig. \ref{fig:highz2}, which also show a general agreement with the $z=0$ HOD shapes. The AGN fractions again peak at $\log~M_{vir}=12.5-13$ [$h^{-1}M_{\odot}$], consistent with previous AGN clustering measurements that show that AGN are typical found in these halo mass scales across a wide range of redshift \citep{Cappelluti:2012,Powell:2020}. However we note that the decline in AGN fraction at $M_{vir}>10^{13}~h^{-1}M_{\odot}$ is less dramatic than at $z=0$, indicating that the $\alpha$ bias may still be present at these higher redshifts, but less severe.

\begin{figure}[h]
    \centering
    \includegraphics[width=.47\textwidth]{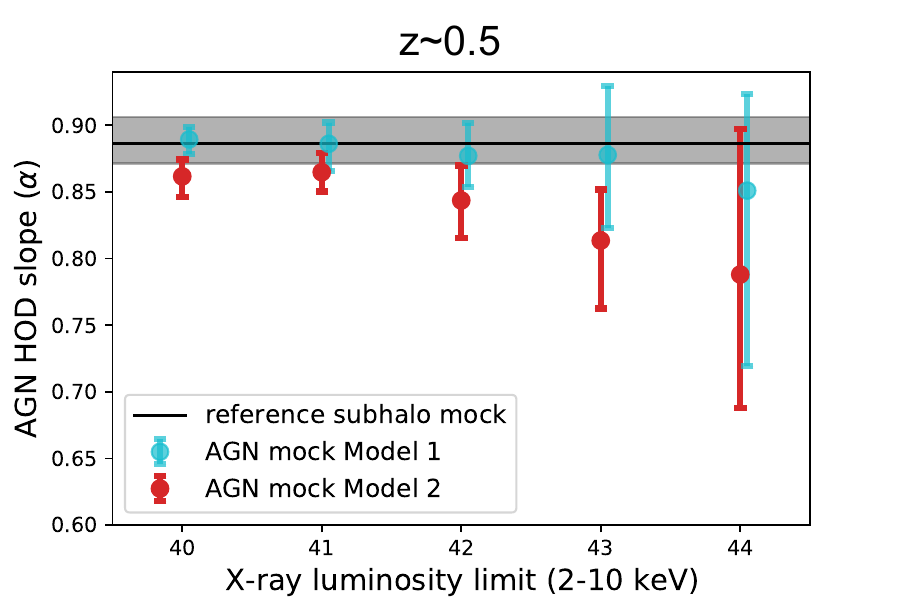}
    \includegraphics[width=.47\textwidth]{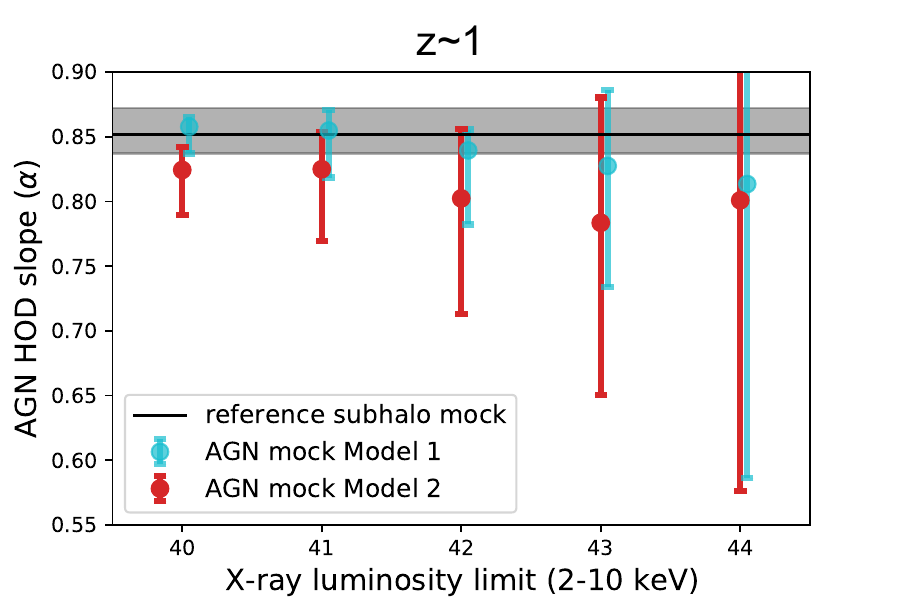}
    \caption{Satellite slopes ($\alpha$) vs. X-ray luminosity limit in the $z=0.5$ snapshot (top) and $z=1$ snapshot (bottom) assuming a universal ERDF. The trends are consistent with those found at low redshift for both Model 1 (cyan) and Model 2 (red); however, the error bars are larger for increasing redshift.}
    \label{fig:highz}
\end{figure}

\begin{figure}[h]
    \centering
    \includegraphics[width=.47\textwidth]{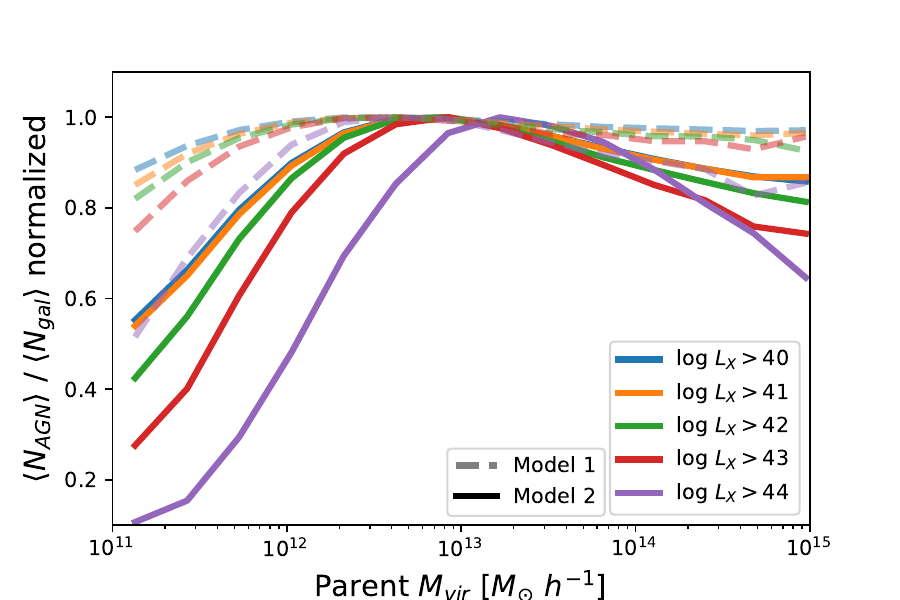}
    \includegraphics[width=.47\textwidth]{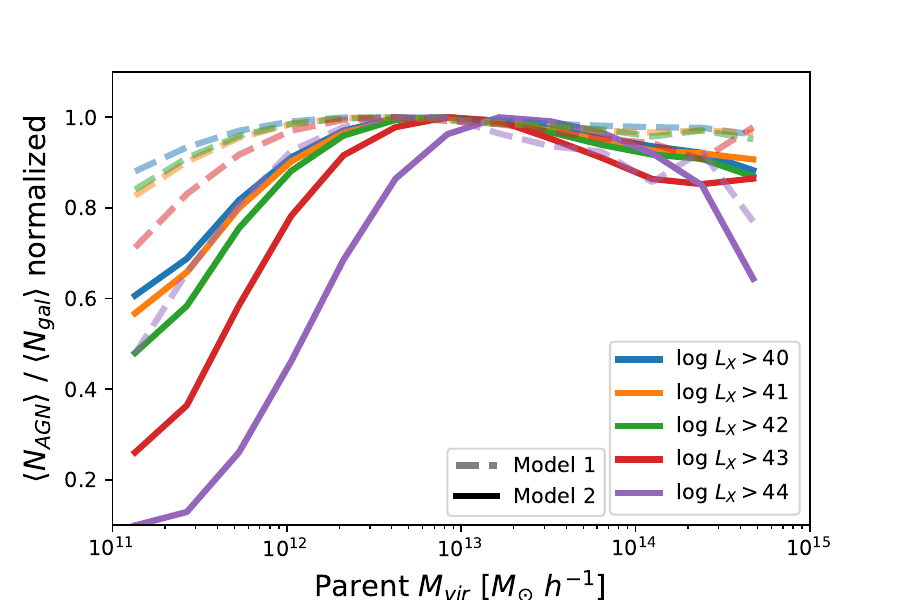}
    \caption{Normalized AGN fractions 
    (averaged over all realizations) vs. parent halo mass for the $z=0.5$ snapshot (top) and $z=1$ snapshot (bottom). Each luminosity threshold is shown in different colors, and the dashed and solid lines correspond to Model 1 and Model 2, respectively. Higher thresholds correspond to AGN fractions that are more peaked, especially for Model 2.}
    \label{fig:highz2}
\end{figure}

\end{appendix}

\end{document}